\documentclass[sigconf]{acmart}

\AtBeginDocument{%
  }

\setcopyright{none}

\acmConference[FAccTRec '24]{7th FAccTRec Workshop on Responsible Recommendation}{October 14, 2024}{Bari, Italy}

\begin{document}

\title{Bias Reduction in Social Networks through Agent-Based Simulations}

\author{Nathan Bartley}

\email{nbartley@usc.edu}
\author{Keith Burghardt}
\author{Kristina Lerman}
\affiliation{%
  \institution{Information Sciences Institute}
  \city{Marina Del Rey}
  \state{California}
  \country{USA}
}

\renewcommand{\shortauthors}{N. Bartley et al.}

\begin{abstract}
Online social networks use recommender systems to suggest relevant information to their users in the form of personalized timelines. Studying how these systems expose people to information at scale is difficult to do as one cannot assume each user is subject to the same timeline condition and building appropriate evaluation infrastructure is costly. We show that a simple agent-based model where users have fixed preferences affords us the ability to compare different recommender systems (and thus different personalized timelines) in their ability to skew users' perception of their network. Importantly, we show that a simple greedy algorithm that constructs a feed based on network properties reduces such perception biases comparable to a random feed. This underscores the influence network structure has in determining the effectiveness of recommender systems in the social network context and offers a tool for mitigating perception biases through algorithmic feed construction. 

\end{abstract}


\maketitle

\section{Introduction}
Recommender systems are pervasive in online social media platforms and they span various functions including social link recommendations (e.g., "Who to Follow"), ad targeting (i.e., users are recommended for ad providers), geolocated news (e.g., Trending Topics), and content recommendation more broadly. While these recommender systems afford users the ability to sift through incredible amounts of information online, they have become the objects of study and critique for their plausible yet not well understood role in amplifying information\cite{ribeiro2020auditing,hussein2020measuring}.

To tease out the impact of the recommender systems, it is important not to overlook the role users play in their interactions with them. Recent work has explored user preferences in agent-based models on YouTube in regards to their primary video recommender system: however an important limitation in this line of work is the lack of comparison of different systems and different kinds of feeds, like those that appear in online social networks like X and Facebook \cite{ribeiro2023amplification,donkers2021dual}. There is a vein of work on X and Facebook that does consider user interactions as a factor in the differences between the black-box algorithmically personalized and reverse chronological feeds, however these works focus more on \textit{what content} is being shown to users rather than \textit{who} the users are being exposed to \cite{gonzalez2023asymmetric,guess2023social,bandy2021more}. Social cues (e.g., the counts of likes and retweets a post has on Twitter/X) and the social context in which people share information (e.g., who they consider their audiences to be) have been shown to impact sharing behavior of posts on social media,  and thus also impacts how information spreads \cite{epstein2022many,marwick2011tweet}. 
To understand how different recommender systems shape users' perceptions of their network, we simulate a Twitter/X-like environment and measure the users' perceived prevalence of a binary trait in the network. With this we can consider how different kinds of feeds skew the perceived prevalence relative to the actual prevalence, which would indicate a bias in how users are exposed to their network.

In this paper we make the following contributions: \begin{enumerate}
    \item We present scaleable agent-based model simulations with $173,000$ nodes.
    \item We compare baselines, two deep-learning approaches and a novel greedy algorithm for personalizing news feeds and measure the exposure bias they generate.
    \item We demonstrate that this greedy algorithm creates less biased feeds and makes feeds that are comparable in utility to the other tested models.
\end{enumerate} 

\section{Related Work}

This work can be categorized under social network simulations, recommender system analysis, and recommender system auditing, as we establish a framework that can be used to plug in and analyze how different implementations of recommender systems in personalized feeds work. We also describe relevant psychological studies and social network phenomena relevant to cognitive biases and perception. 

\subsection{Social Network Simulations}

Social network multi-agent simulations for recommender system analysis have been largely focused on information diffusion and predicting user behavior in different environmental circumstances. Muric et al.  use Twitter, Reddit and GitHub data to understand information spread, especially across different online platforms \cite{muric2022large}, whereas we explicitly focus on comparing different personalized feeds within a platform .

Murdock et al. \cite{murdock2023agent} simulated user and moderation behavior on Reddit, which is an interesting additional social layer to consider. In our work we do not consider moderation in an effort to simplify our assumptions about user behavior. 

Ribeiro, Veselovsky, and West \cite{ribeiro2023amplification} utilized an agent-based model to address the paradox that content-based recommender systems face: these systems do not seem to be the primary driver of what users consume even though they favor extreme content. Their results suggest users will not engage with suggested low-utility content. Our work differs in that we compare different recommender systems and how they would behave considering the same user behavior patterns. Our work also considers platforms that use more social information in the recommendations, as a piece of content might appear in your feed if your friend interacts with it. 

Donkers et al. \cite{donkers2021dual} studied both epistemic and ideological echo chambers in social media and the effects of recommendations on depolarizing discussions. While depolarizing discussions is an important goal in this line of work, in our current study we simplify the models and compare their effect on perception instead.

\subsection{Recommender System Audits}

In general recommender system audits have focused on content-based recommender systems as they are the most straightforward to analyze. In particular, most recent work has focused on YouTube and the role the video recommendation engine might have in spreading misinformation and radicalization. These works, like Tomlein et al. \cite{tomlein2021audit} and Hussein et al. \cite{hussein2020measuring} used agent-based sock puppets to simulate user behavior directly on the platform. This contrasts to our work as we simulate the platform as well.

Both Spinelli and Crovella \cite{spinelli2020youtube} and Ribeiro et al. \cite{ribeiro2020auditing} investigated YouTube's video recommender system and how it might contribute to gradual user radicalization. The latter study in particular analyzed user interactions and comments, focusing on migrations of users between communities. Understanding user dynamics are essential, however focusing on YouTube overlooks social signals: the same information may be perceived positively by a user if it was presented as being "liked" by a friend, but negatively by that user if it was presented by a stranger or someone they dislike. This makes such content-based studies about recommendations less relevant for platforms like X and Facebook.

\subsection{Personalized News Feeds}

Most studies investigating the effects of personalized news feeds, defined as the ordered information a user receives when logging onto a platform like X/Twitter, are focused around user satisfaction, impact on information diffusion, and echo chambers.

Two papers addressing perception and user satisfaction are Eslami et al. \cite{eslami2015feedvis} and Eslami et al. \cite{eslami2016first}, where tools were developed to study how users perceive the Facebook news feed algorithm at that time, i.e., they studied the impact of algorithmic awareness on user satisfaction and perception of their networks.

Bakshy et al. \cite{bakshy2015exposure} from Facebook addressed information diffusion via the personalized News Feed by measuring user exposure to ideologically diverse posts. In this they considered how user preferences and algorithmic influence play a role in content consumption. They focused on the dynamics of information diffusion in Facebook's network at the time and did not emphasize the effect of different personalized news feeds.

More recently, Guess et al. \cite{guess2023social} investigated the Facebook and Instagram feed algorithms and their potential impact on user attitude and election behavior in the 2020 presidential election. Gonzalez-Bailon et al. \cite{gonzalez2023asymmetric} studied the interaction of algorithmic curation and user social amplification by studying the spread of political URLs on Facebook, showing a segregated experience between liberal and conservative users. These papers are relevant to studying social network based recommendation systems, however they are focused on political content diffusion and political-behavior related outcomes for real users: our current study is concerned with comparing different types of personalized feeds and their impact on user exposure. This exposure is important as perception of the prevalence of traits in a network can potentially impact beliefs and behaviors related to those traits. 

\subsection{Psychological \& Network Phenomena}
Perception biases can be connected to how we perceive our social environment, but they can also be tied to various psychological phenomena like the illusory truth effect \cite{hassan2021effects} and the mere exposure effect \cite{zajonc1968attitudinal}. These two effects, which describe an individual's assessment of a stimulus after multiple repeated exposures to it, suggest that a user of a social media platform may be influenced by how often they observe specific user traits or opinions online. 

As it pertains to social stimuli we also address the salience bias as described by Kardosh et al.: people across multiple cultural contexts were more attentive to unexpected or irregular stimuli, which in this case was the fraction of faces belonging to minority groups in a visual matrix of faces\cite{kardosh2022minority}. Participants in both the minority and majority groups studied systematically over-estimated the prevalence of the minority faces in a recall task. This suggests that how often you observe a trait in a social media feed may impact your perception of how prevalent it is.  

These cognitive biases are important as there are well understood structural phenomena in social networks that impact individuals' local perception of global characteristics (e.g., the majority illusion and the friendship paradox) \cite{lerman2016majority}. With phenomena like the majority illusion, people in a social network are likely to perceive a trait as being more common than it truly is. 

\section{Model}
\subsection{Framework}

We use the Repast framework to run models over a cluster of compute nodes \cite{collier2011repast}. This framework has previously been used in other areas like simulating bike-sharing systems in cities, and complex biological systems requiring heterogeneous multicellular organisms \cite{montagud2021systems,bae2024data}. A key factor that aided our use of Repast is that in our simplified network, users will only be exposed to the tweets that their friends (and friends-of-friends) generate, which allows us to partition the network and run them in different processes. This ability to partition the network allows for significant scalability. 

In this framework we model an edge as a connection between two users. Each edge represents a follow connection, and users are exposed to tweets generated by users they are connected to (friends) or by users their friends are connected to (friends-of-friends). Edges are observed if they are presented to that user (here if a friend-of-friend is observed a \textit{de facto} edge is observed).

\subsection{User Behavior}
In this work we trace the behavior of approximately 173,000 users (each of whom is labeled $x=1$ with a fixed probability $P(X=1)$ otherwise $x=0$) sharing 1.5 million edges. This trait $x$ can represent a demographic trait or an opinion that is held by each user. With this trait we then examine the experience of 5,599 central users as they follow the below sequence of events for each time tick $t$: \begin{enumerate}
    \item Activate user $i$ with a uniform probability (0.083)
    \item If activated, user $i$ then produces a certain number of tweets; we sample a lognormal distribution ($\mu = 0.0, \sigma = 1.0$) to choose how many tweets the user produces in that time period
    \item Add created tweets to the content pool
    \item "Backend" model serves tweets to user $i$ if appropriate
    \item User $i$ likes tweet from another user with same label at fixed probability (0.20), with probability (0.05) otherwise
\end{enumerate}
Our model is a discrete event-based model, where for each time tick we "step" the individual nodes and then update model information before proceeding to the next time tick.

\begin{figure*}[ht]
\centering
\includegraphics[width=0.75\textwidth]{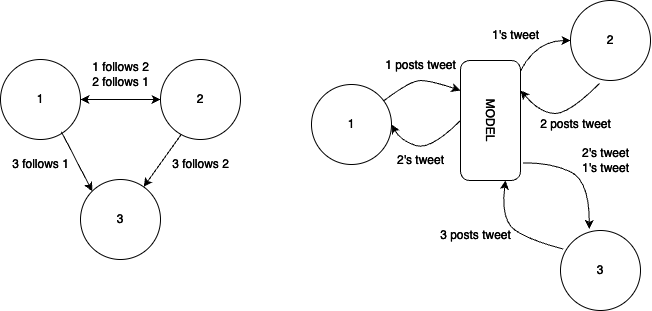}
\caption{\textbf{Agent-based Model Structure} Illustration demonstrating how three users are connected to each other on the network, but will only get exposed to other users through the tweets served to them from the "backend" model. }  \label{fig:model}
\end{figure*}

There are two key components for the model: the  "backend" model that serves users tweets and the network of users. While each user is connected via the network, they only interact with other users on the network through the model  by sending tweets to the model and getting tweets from their friends through the model. This way we can vary how tweets are served to users. We illustrate how this model works visually in Fig. \ref{fig:model}.

At tick $t=24$ we reset the edges seen by users in the network to reflect a full 24 hours passing, in order to assess what happens when the network "forgets" most information from the previous day. This is also a validity check to ensure that the consistency in the dynamics of the system (i.e., we want to make sure that the system will converge back to a similar point as before the "reset"). 

We structure the network based on data from Alipourfard et al., 2020 \cite{alipourfard2020friendship} who gathered a complete follower network for $5,599$ users, as well as tweets and retweets for those users and everyone they followed to generate a dataset with 4M users from May - September 2014. We use this data to guide our model; we downsample the nodes for the sake of simulation runtime.

\subsection{Model Parameters}

We treat each simulation time step as a single hour, for a total of 36 timesteps. Per an official blog-post from X \cite{twittertime}, users spend an average of 32 minutes per day on the platform, which we implement in an activation probability: each user has a 0.083 probability of "logging in" per hour to give an expected value of approximately two sessions per 24 hours.

To assess perception of networks, we randomly assign each user in the network a binary trait $X \in \{0,1\}$ such that the total prevalence of the trait in the network is static. We run each simulation under $P(X=1) \in \{0.05, 0.15, 0.50\}$ to assess the impact of the prevalence of the trait on the behavior of the system. Each user, based on the value of the trait that they are assigned, also behaves in a biased manner towards the tweets that they observe: users with $x=1$ will like tweets from users with $x=1$ with probability $0.20$ and will like any other tweets with probability $0.05$. Likewise for users with $x=0$. Both numbers were chosen to elicit, in expectation, at least one like from each active user per tick. 

Assigning traits to the nodes allows us to measure the degree-attribute correlation $\rho_{kx}$, which is defined as:
\begin{equation}\label{eqnrho}
    \rho_{kx} = \frac{P(x=1)}{\sigma_{x}\sigma_{k}}[\langle k \rangle_{x=1} - \langle k \rangle]
\end{equation}

Here $\langle k \rangle_{x=1}$ is the average in-degree of the "active" $x=1$ nodes, $\langle k \rangle$ is the average in-degree of all nodes considered, and $\sigma$ represents the respective standard deviation.

Each simulation has every user subjected to the same personalized news feed: \begin{enumerate}
    \item \textbf{Random.} All candidate tweets are randomly sorted and the first $n$ tweets are served to the user.
    \item \textbf{Reverse Chronological.} All candidate tweets are sorted in reverse chronological order, and the first $n$ tweets are served to the user.
    \item \textbf{Neural Collaborative Filtering.} We implement a basic version of the Neural Collaborative Filtering (NCF) model to showcase how deep learning-based recommender systems operate in a social network context. We want to demonstrate what happens when the model has separate user and item embeddings and the ability to capture latent user-item interactions. We train the model on the 5,599 core users, where each tweet liked is the "item" being trained on. We keep the model straightforward and only use the superficial user level information \cite{he2017neural}. Model is trained every tick for 10 epochs and under a binary focal cross-entropy loss function.
    \item \textbf{Wide \& Deep.} We implement a simple version of the Wide \& Deep model described initially by researchers at Google to demonstrate how a recommender system used in production in other contexts might behave in this scenario \cite{cheng2016wide}. We use the same features as the NCF model for training. Model is trained every tick for 10 epochs and under a binary focal cross-entropy loss function.
    \item \textbf{Minimize $\mathbf{\rho_{kx}}$.} We implement a greedy strategy for choosing which edges to observe for each user. We use eqn. \ref{eqnrho} and sort the tweets seen by a user at every tick $t$ by how much that edge would contribute to the difference between the mean "active" in-degree $\langle k \rangle_{x=1}$ and mean in-degree $\langle k \rangle$, opting for the edge that would minimize the difference.
\end{enumerate}

We chose these personalized news feed algorithms to analyze different implementations of news feeds: there is a public release of the X recommendation "algorithm", however as it relies on multiple models and active services we cannot use the code as-is in a simulation (especially as production model parameters have since changed) \cite{twitteralgo}. Instead we aim to show simple baseline models, two deep-learning models, NCF and Wide \& Deep, and our greedy strategy for minimizing exposure bias (MinimizeRho).

Each simulation similarly has every user using the same length feed, i.e., they only observe (and potentially engage) a fixed number of tweets for each timestep in the simulation. We tested lengths of 30, 50, and 100. 

\subsection{Bias \& Performance Metrics}

To operationalize the exposure bias we are studying here we rely on the social network structural phenomena described previously. Here \textit{exposure bias} refers to the over- or under- representation of a trait in a user's feed, measuring the trait's perceived prevalence in the network relative to its true prevalence. 

We use two metrics to study the bias of this perception:
\begin{enumerate}
    \item Local bias ($B_{\text{local}}$) \qquad \qquad $B_{\text{local}} = E[q_{f}(\text{X})] - E[f(\text{X})]$ 
    \item Gini coefficient (G) \qquad \qquad $G = \frac{\sum_{i=1}^{n} (2i - n - 1)x_{i}}{n\sum_{i=1}^{n}x_{i}}$ 
\end{enumerate}

We define $B_{\text{local}}$ as the average frequency of the attribute among a node's immediate network: $E[q_{f}(\text{X})] = \bar{d} *  E[f(U)A(V)|(U,V) \sim \text{Uniform}(E)]$; $E[f(X)]$ is the global frequency of the node attribute f (here, 0.05, 0.15, 0.50);  $f(U)$ the attribute value $f$ of node $U$; $\bar{d}$ represents the expected in-degree of the network. $A(V)$ represents the ``attention`` node $V$ pays to any node in their network. $B_{\text{local}}$ should vary from [-1,1], and Gini should vary from [0,1], where 0 is equal and 1 is unequal. This plays the role of an overall view of the skew users will experience in their personalized feeds. 

Gini coefficient in this context represents the skew in the number of unique friends (or friends-of-friends) users are exposed to in their feeds: $x_{i}$ represents the number of times that friend (or friend-of-friend) was observed.

We use several other measurements to study the simulation and verify results (some not reported here due to space constraints):
\begin{enumerate}
    \item \textbf{Precision@10.} $\text{mean}\left(\frac{|\text{cumulative tweets liked in first 10 positions}|}{10}\right)$
    \item \textbf{Precision@30.} $\text{mean}\left(\frac{|\text{cumulative tweets liked in first 30 positions}|}{30}\right)$
    \item \textbf{Number of edges seen.} Total unique edges seen up until that time tick, including friends-of-friends. 
    \item \textbf{Mean number of likes friends' tweets receive.}  We take the sum total likes each friend receives from core users and take the mean over all friends.
    \item \textbf{Mean number of likes given.} Mean number of likes given by the 5,599 core users over the course of the simulation.
\end{enumerate}

\section{Results}
\begin{figure*}[ht]

\includegraphics[width=0.45\textwidth]{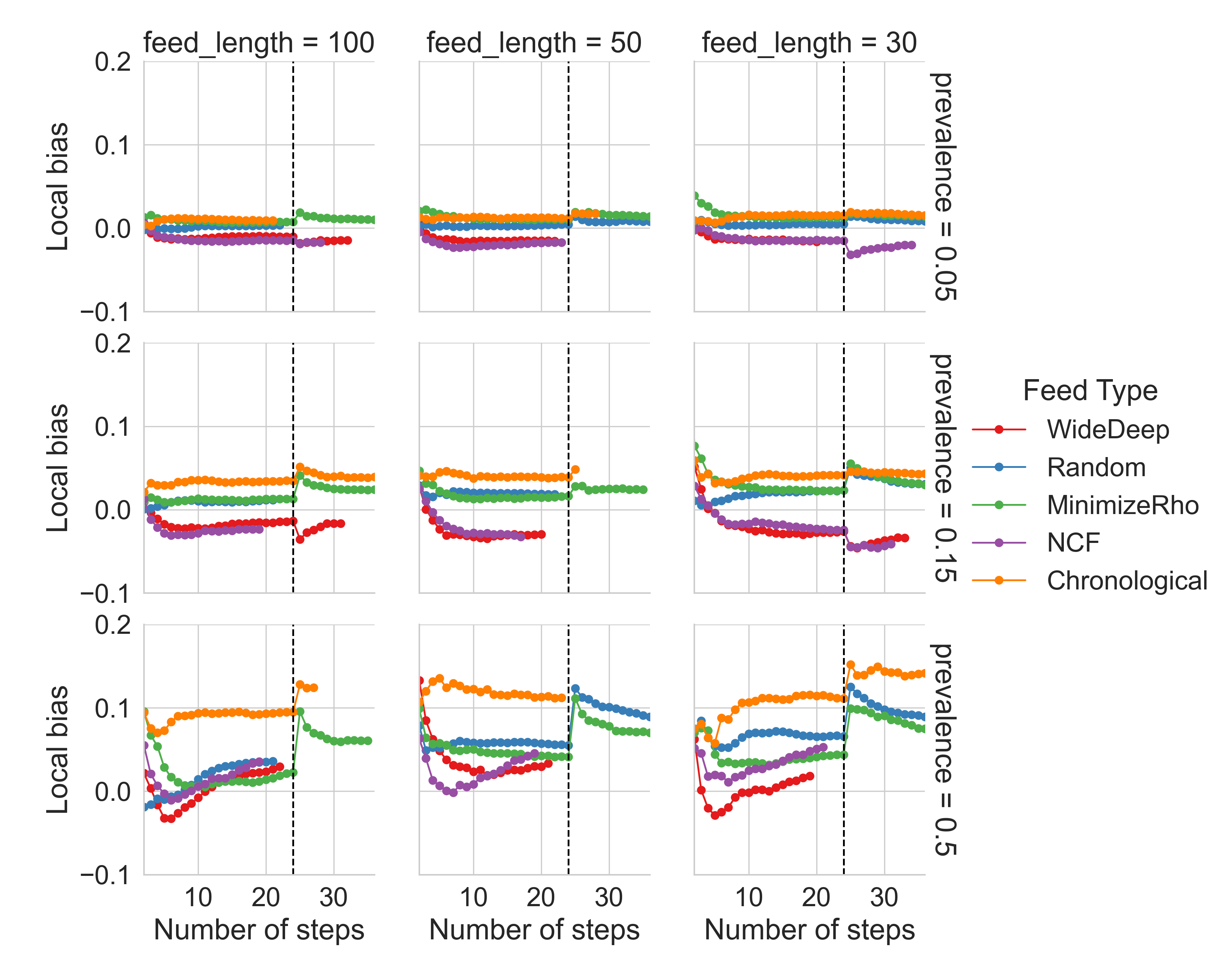}
\includegraphics[width=0.45\textwidth]{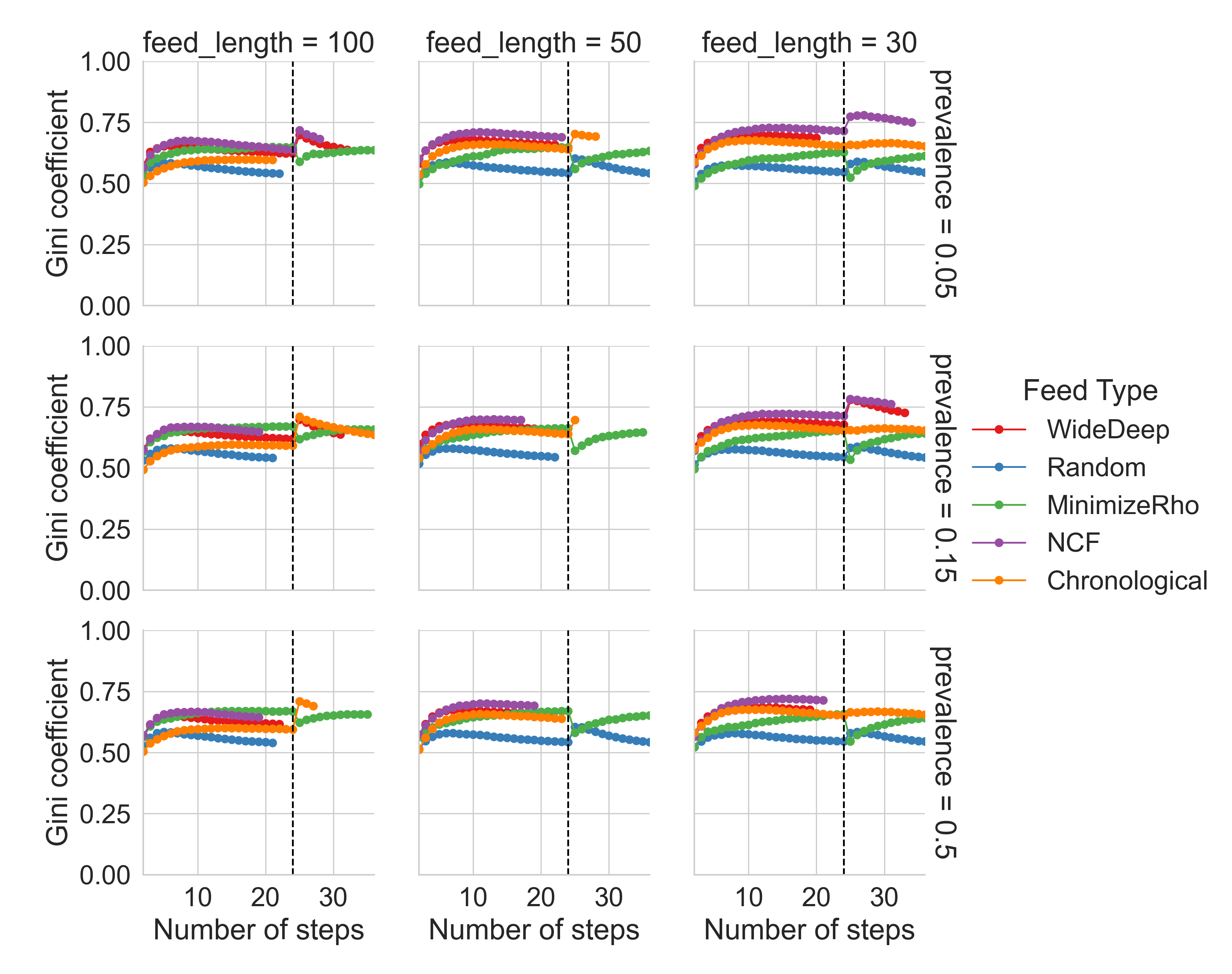}
\caption{\textbf{Local bias ($B_{\text{local}}$) and Gini coefficient G.} Graph depicts the difference between the expected local fraction of friends who have $x=1$ and the true global prevalence of the trait $P(X=1)$. Positive implies over-representation, negative implies under-representation. For G, graph depicts the distribution of times each friend (or friend-of-friend) was observed by a core user. 1 implies inequality, 0 implies equality. }\label{fig:whole_bias}
\end{figure*}

Across models in Fig. \ref{fig:whole_bias} we observe relatively stable bias for the network until the reset at $t=24$. Interestingly, we observe that the Random and MinimizeRho conditions have consistently low measures (in absolute value) of $B_{\text{local}}$ (Local Bias). Similarly, the NCF and WideDeep conditions are correlated to one another, showing negative values (i.e., they under-expose the users to users with trait $X=1$).

In Fig. \ref{fig:whole_bias} we observe converging, stable behavior of the Gini coefficient, where the Random and Chronological conditions remain the lowest in terms of Gini, suggesting a more even distribution of attention across friends. Interestingly the MinimizeRho condition starts low and progresses higher to become similar to the Chronological, NCF, and WideDeep conditions. These two NCF and WideDeep models tend to have the highest Gini coefficient, suggesting a narrower focus on sets of friends observed by users. This remains consistent even after the $t=24$ reset where the MinimizeRho condition again starts low and progresses higher in Gini.

\begin{figure*}[]

\includegraphics[width=0.45\textwidth]{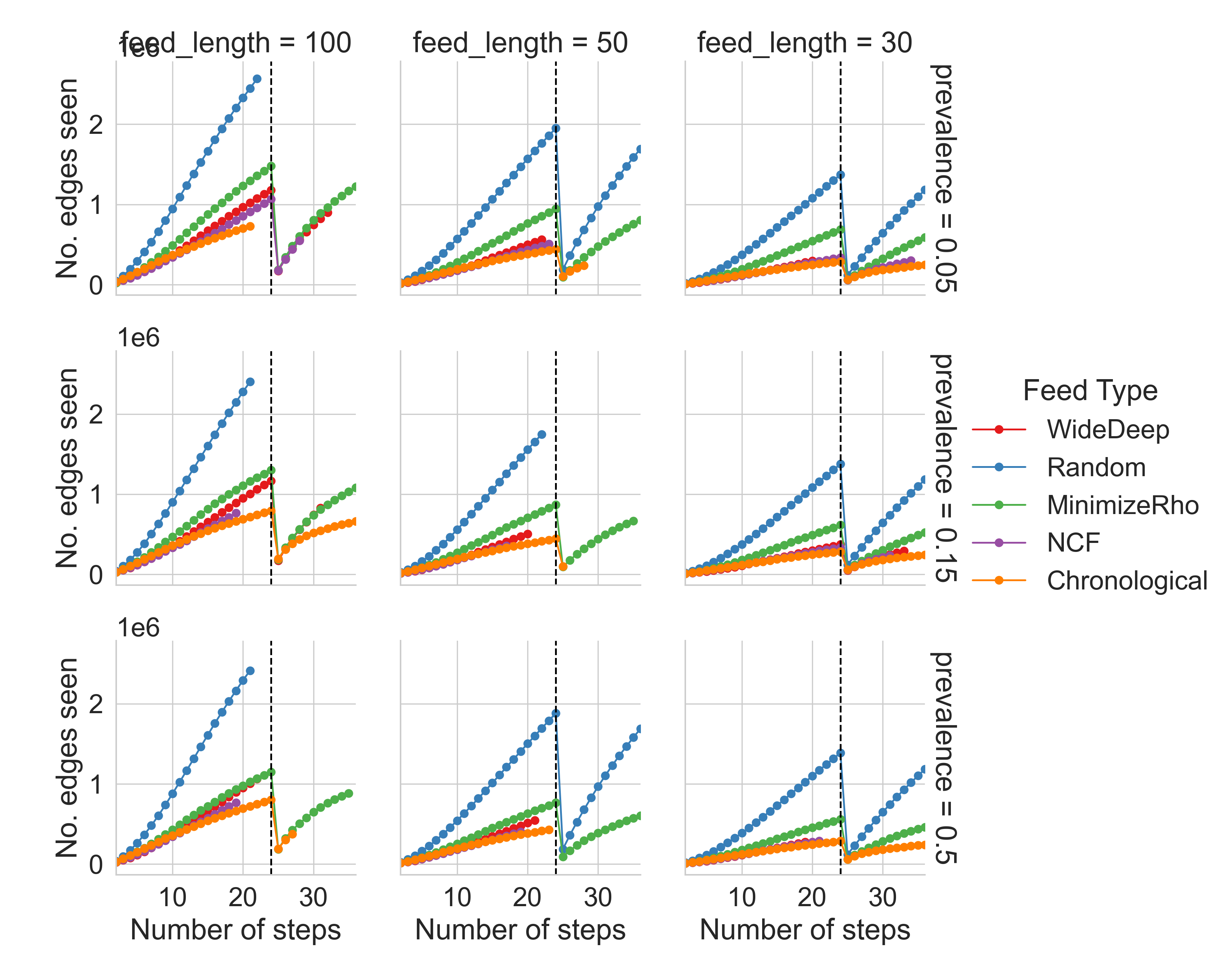}
\includegraphics[width=0.45\textwidth]{plots/precision_at_30nochg6.png}
\caption{\textbf{Log number of friends seen and Precision@30}. Graph depicts the log total number of unique friends (and friends-of-friends) seen through tick $t$. Connections seen are reset at $t=24$. For Precision @30, graph depicts the total fraction of liked tweets in the first 30 positions in the feed}\label{fig:num_edges_seen}
\end{figure*}

For validity checks of the simulations we present the results in Fig. \ref{fig:num_edges_seen}. Here we observe the unique edges observed by the core users, where the Random condition maximizes the total observed edges, followed by the MinimizeRho condition. The number of likes given and received in Figs. \ref{fig:num_likes_gen} and \ref{fig:num_likes_rec} shows that the 5,599 core users all behave similarly under different conditions in terms of the number of tweets that they like  cumulatively over the course of the simulation. The longer feed length simulations tend to have higher mean likes than the shorter ones. We observe similar behavior in the mean number of likes each friend receives, with longer feeds having higher mean tweets cumulatively over time.

\begin{figure*}[]

\includegraphics[width=0.45\textwidth]{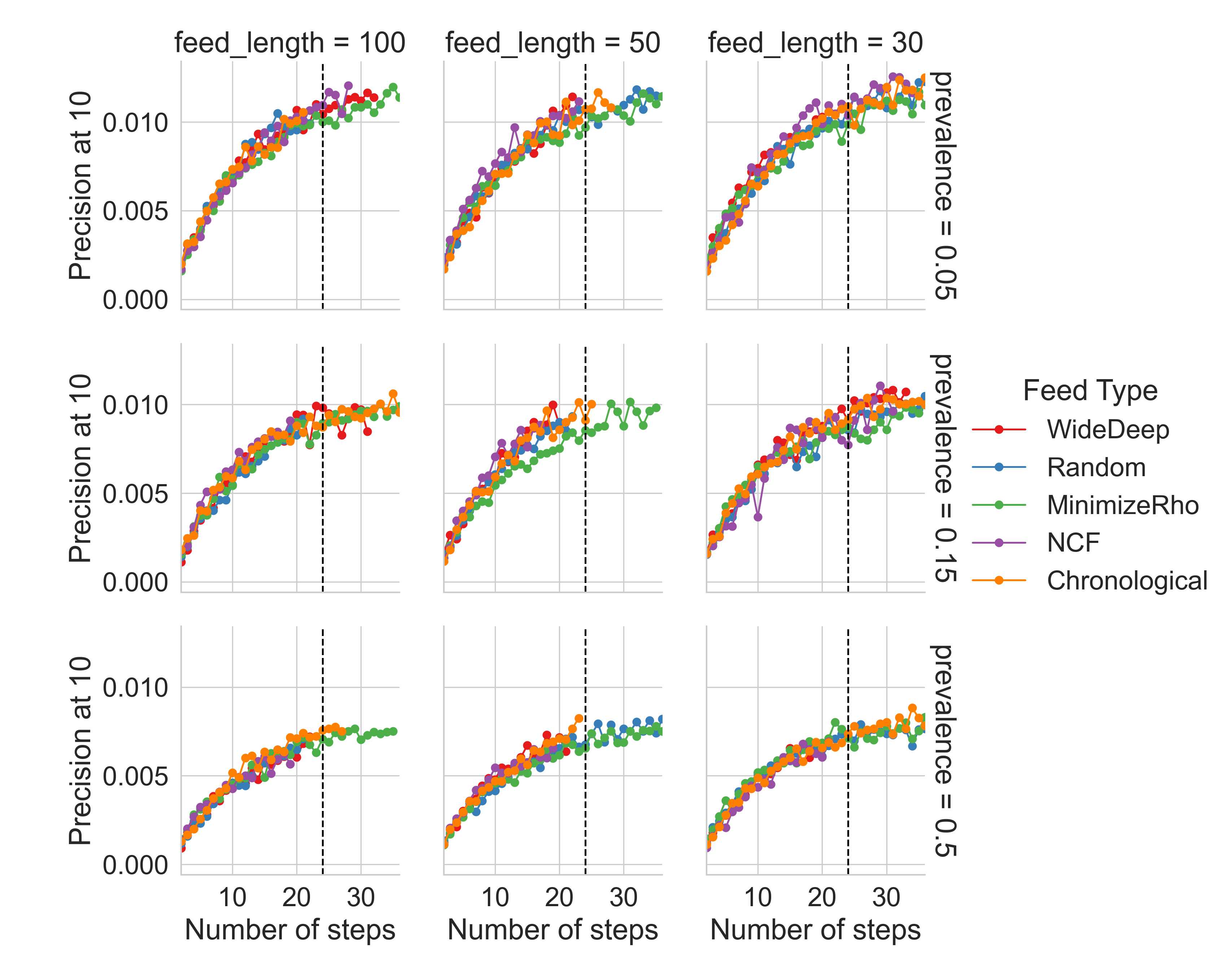}
\includegraphics[width=0.45\textwidth]{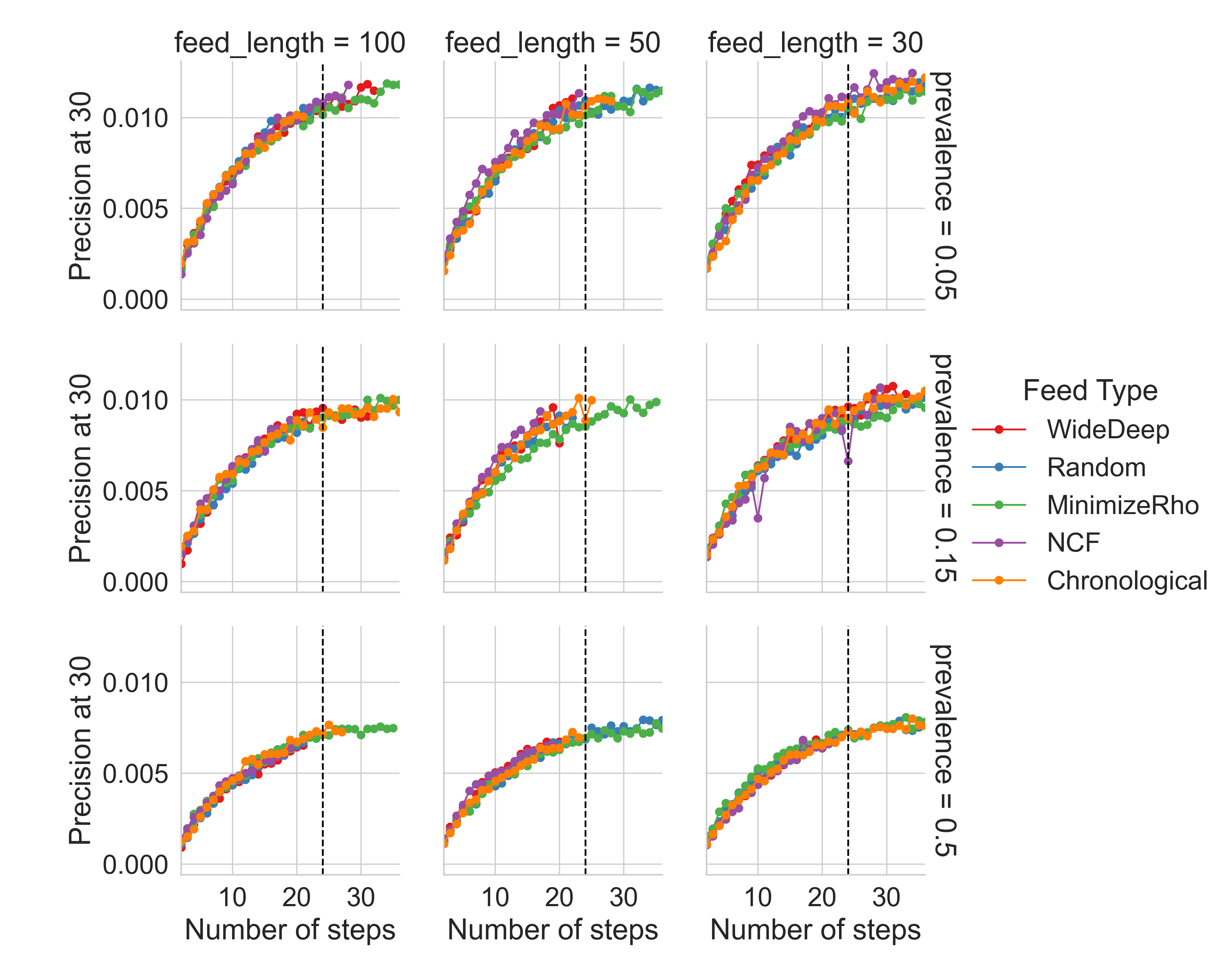}
\caption{\textbf{Precision@10 and Precision@30}. Graph depicts the cumulative number of tweets liked in the first K positions seen through tick $t$. Connections seen are reset at $t=24$. For Precision @30, graph depicts the total fraction of liked tweets in the first 30 positions in the feed.}\label{fig:precision}
\end{figure*}

The precision figure in Fig. \ref{fig:precision} demonstrates that all model conditions improve over time, with different levels of minor precision drop after the $t=24$ reset. The precision is computed cumulatively over the course of the simulation.

\section{Discussion}

The information presented in Fig. \ref{fig:num_edges_seen} suggests that the Random and MinimizeRho conditions demonstrate the most growth in number of unique edges observed over time relative to the other models, increasing diversity in users who are observed. As this might be due to spurious changes in user behavior, we measure the mean number of likes each friend receives in Fig. \ref{fig:num_likes_rec}. Here we observe that the longer feeds provides more likes for each friend, which follows given the increased number of "chances" each user would get to like a friend's tweet. Given each model has users that behave similarly, we then use the perception measures in Fig. \ref{fig:whole_bias} to determine differences between models. Two patterns show up: the Random and MinimizeRho conditions correlate closely and lower in absolute value in $B_{\text{local}}$, whereas the NCF and WideDeep models behave tightly, but with lower (negative) values in $B_{\text{local}}$. This suggests that the deep learning models are more tightly focused on certain sets of users (as corroborated in the number of edges seen in Fig. \ref{fig:num_edges_seen}). Interestingly, the MinimizeRho condition behaves similarly to the Random condition in Fig. \ref{fig:whole_bias} in the initial timesteps, however it diverges and becomes more similar to the Chronological condition in most conditions (it becomes more like the deep learning models in the feed length of 100). This is interesting as it seems to be more sensitive to the feed length and prevalence than the other models tested in terms of Gini. This suggests that the MinimizeRho model converges to a similarly focused set of users as the deep learning models, but because it presents a comparatively undistorted view of the network the MinimzeRho model may be ignoring edges that might offer utility to users but skew the view of the network. If the Random model can be considered less biased than the others by minimizing the absolute value, then the MinimizeRho model seems to be most closely related to it in behavior over time (albeit in a more focused deterministic manner than the Random condition). 

Curiously, each model trends upwards in performance in Precision@30 in Fig. \ref{fig:num_edges_seen}. We would expect the Random model to have minimal slope as it does not depend on the user feedback, however, as the number of likes is stored cumulatively we would anticipate a growing number over time. This behavior may reflect previous work in understanding structural biases in networks: Lee et al., 2019 demonstrate that the homophilic structure of a network and the size of a minority group can impact similar perception biases \cite{lee2019homophily}. In other words, the effects of the different recommender systems may be marginal to the effect of how the network is constructed and the distribution of the trait, as reflected in the different observed slopes across the prevalences in Fig. \ref{fig:num_edges_seen}.

To describe these differences more we observe that under $P(X=1)=0.05$ the NCF model is one of the better models in both measures of precision. However the difference in models disappears as the prevalence of the trait tends towards $P(X=1)=0.50$. The drop in performance of the MinimizeRho model at feed length 50, $P(X=1)=0.15$ may be explained by the static assignments of the trait $X$ for those simulations: other versions of these simulations where we modulate the correlation $\rho_{kx}$ seems to remove this visible discrepancy.

Overall, while behavior of the simulated system appears to converge, it is unlikely that the real ecosystems being modeled would converge so readily. If we do assume some stability it seems to be the case that personalization would lead some users to perceive that any particular trait is more (or less) prevalent in their larger social networks than they actually are. In other words, it seems that personalization can either mitigate or amplify network-based structural phenomena like the majority illusion \cite{lerman2016majority}.

\section{Limitations \& Future Work}

One clear limitation in our study is the duration for which we ran the simulations. Some of the more complex recommender feeds (and longer length feeds) took longer than expected to run. This is another limitation: we do not try more recommender systems and longer personalized feeds. There are  advanced recommender systems that could be useful to analyze, like MV-DNN \cite{elkahky2015multi} or the stated X/Twitter system \cite{twitteralgo}, however adapting such models is difficult as many require access to richer information about the users and content than we have built here (or for a production system we would need access to other microservices or models the system depends on). This complication aside, these more complex systems would afford us the ability to compare the results of these ABM studies more directly to real user studies, allowing us to tune the ABM parameters to be more accurate to real user behavior. However, it behooves us to be wary of running ABMs that are too contrived to be useful, as they can be difficult to reproduce and apply elsewhere.

As some of the results may possibly be explained by confounding factors from the network itself, i.e., the structure and how the trait is distributed, future work should entail simulations on multiple kinds of networks. 

Using Large Language Models as agents interacting in the framework would be interesting future work, considering Tornberg et al. \cite{tornberg2023simulating} and their preliminary work in this space. This could facilitate the use of complicated recommender systems, as other ways of generating richer information  would again potentially make the ABM too contrived to be useful, as described above. 

We would also like to integrate more metrics into the analysis, as there may be confounding factors present in our simulations (e.g., $B_{\text{local}}$ and $G$ may converge but some other measure might be periodic). To analyze how different groups of users experience the exposure bias, we would like to examine user subgroups (either by network structure or by label). Similarly we would like to extend this analysis to more than binary user labels as labels can change over time and are often more complicated than simple binaries. In an effort for reproducibility we release the simulation scripts.\footnote{\url{https://github.com/bartleyn/cuddly-octo-couscous}}

\section{Conclusion}

In this work we describe an agent-based model and framework for studying the effects of different personalized news feed algorithms in online social networks by measuring how they expose users to their networks. The model and framework is extensible and given the MPI usage of the underlying Repast library very scalable contingent upon having access to an MPI-enabled computing resource. More complex user behaviors are straightforward to implement, and additional models can be implemented as well for the underlying recommender system. We find that while deep learning methods are useful and tend to minimize perception bias in terms of our binary label, they focus on a narrower set of users. Our findings show that a simple greedy algorithm, which selects content based on network properties, increases diversity in the attention users give to their network. This algorithm also minimizes the absolute value of local perception bias. This suggests that platforms should consider how traits are distributed across the network as a feature for the various recommender systems serving users' timelines. 

These findings are important for designing recommender systems in online social networks that are robust: these systems mediate the information and connections between people and we should be able to understand what happens as people interact with these dynamic and ubiquitious systems.

\section{Ethical statement}

We generally believe that agent-based models are an appropriate method for studying these systems in a way that preserves user privacy and dignity in this area of research. We do have concerns however that the more we understand recommender systems, the more of an attack vector we open up for malicious actors to manipulate these systems. Considering the recent (as of this writing) layoffs at major social media platforms, and a changing focus in Trust and Safety, this could pose a problem for the spread of harassment and misinformation.

\bibliographystyle{ACM-Reference-Format}
\bibliography{manuscript}

\appendix

\begin{figure*}[h]
\centering
\includegraphics[width=0.75\textwidth]{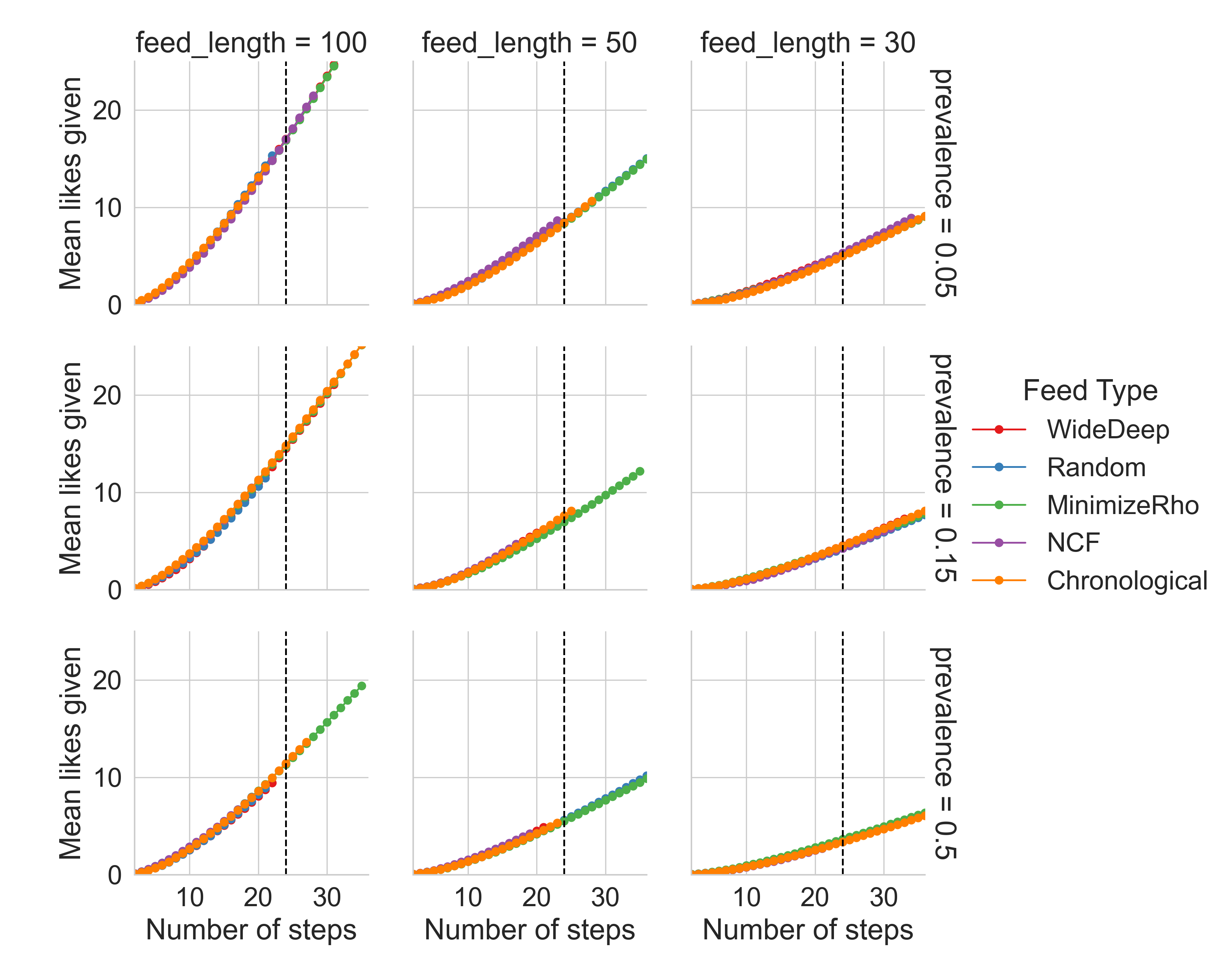}
\caption{\textbf{Mean number likes generated by core users}.}\label{fig:num_likes_gen}
\end{figure*}

\begin{figure*}[!ht]
\centering
\includegraphics[width=0.75\textwidth]{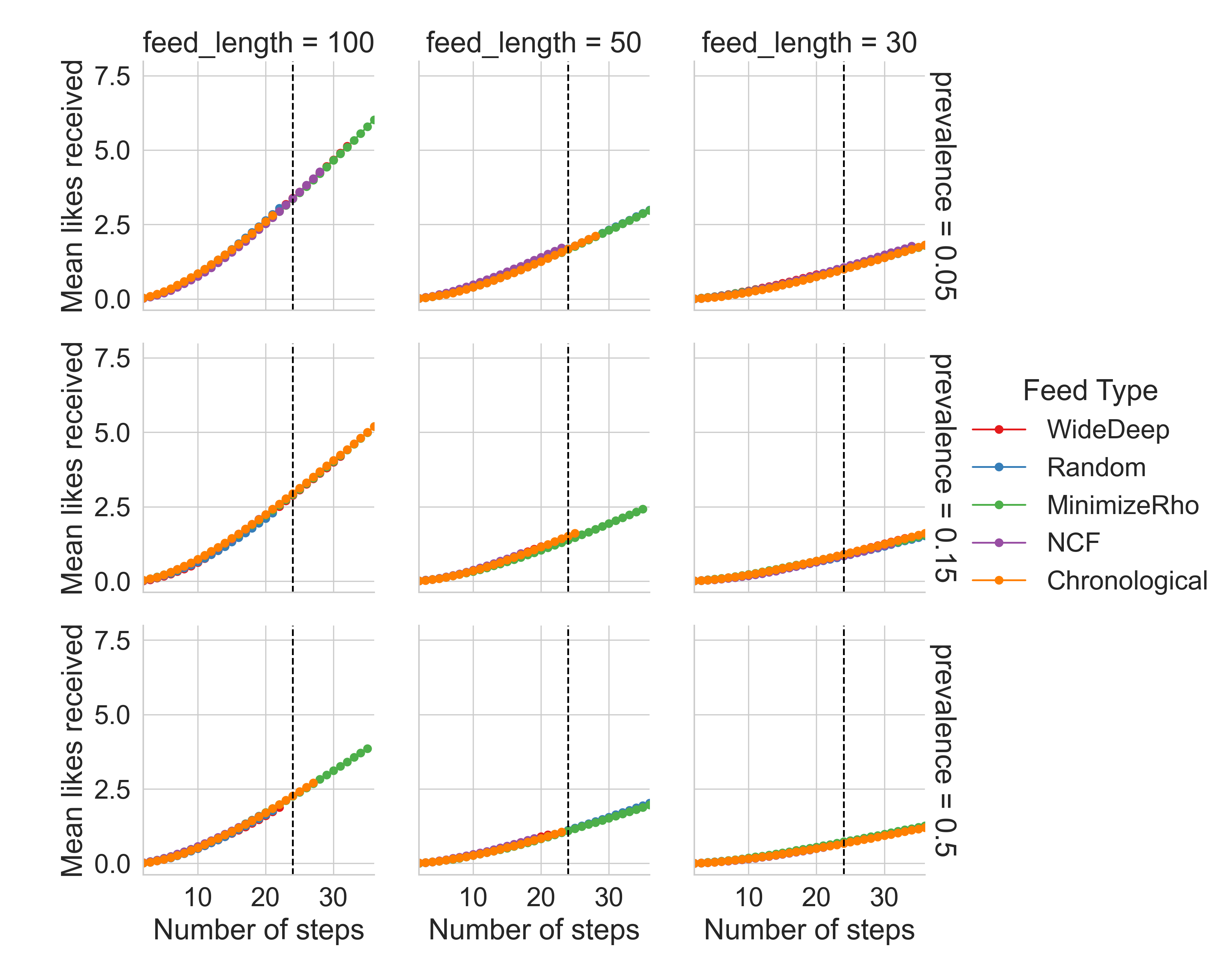}
\caption{\textbf{Mean number of likes each friend receives}. Graph depicts the mean number of likes each friend receives over time. }\label{fig:num_likes_rec}
\end{figure*}

\end{document}